\documentclass[twocolumn,showpacs,preprintnumbers,prl]{revtex4}
\usepackage{eurosym}
\usepackage{graphicx,bm,amsmath,amssymb}

\def\gz{\ifmmode{Z\hskip -4.8pt Z}
    \else{\hbox{$Z\hskip -4.8pt Z$}}\fi}

\newcommand{\be}{\begin{equation}}
\newcommand{\ee}{\end{equation}}
\newcommand{\bea}{\begin{eqnarray}}
\newcommand{\eea}{\end{eqnarray}}

\begin{document}

\title{Leading temperature dependence of the conductance in Kondo-correlated quantum dots}
\author{A.~A.~Aligia}
\affiliation{Centro At\'{o}mico Bariloche and Instituto Balseiro, Comisi\'{o}n Nacional
de Energ\'{\i}a At\'{o}mica, CONICET, 8400 Bariloche, Argentina}

\begin{abstract}
Using renormalized perturbation theory in the Coulomb repulsion, we derive an analytical expression 
for the leading term in the temperature dependence of the conductance through a quantum dot described by the
impurity Anderson model, in terms of the renormalized parameters of the
model. Taking these parameters from the literature, we compare the results
with published ones calculated using the numerical renormalization group
obtaining a very good agreement. The approach is superior to alternative perturbative treatments.
We compare in particular to the results of a simple interpolative perturbation approach.
\end{abstract}

\maketitle


\section{Introduction}

\label{intro}

The manifestations of the Kondo effect in transport through semiconducting 
\cite{gold,cro,gold2,wiel,grobis,keller} and molecular 
\cite{leuen,parks,scott,parks2,serge} quantum dots (QDs) were a subject
of great interest in the last years. The Kondo effect takes place when the
occupancy of the isolated QD is such that its spin $S \neq 0$. For temperatures $T$
below the Kondo temperature $T_K$ this spin is totally or partially
compensated by the conduction electrons of the leads leading to a many-body
ground state with lower total spin. This implies a resonance at the Fermi
energy in the spectral density of the dot state, that leads to an anomalous
peak in the differential conductance $G(V)=dI/dV$ at zero bias voltage $V$,
where $I$ is the current through the QD. For the simplest systems with one
relevant level and $S=1/2$, these physical effects are usually well
described by an impurity Anderson model (IAM), which contains the Kondo
model as the limiting case in which valence fluctuations are absent \cite{hewson}. 
The parameters of the IAM are the
energy of the dot level $E_d$, the resonant level width $\Delta$ and the
Coulomb repulsion $U$. The Kondo regime corresponds to $-E_d, U+E_d \gg
\Delta$ \cite{hewson}.

In the Kondo limit, the properties of the IAM display universality. Physical
observables are described by the same universal function, once the different
physical magnitudes are scaled by $T_K$. For example, it has been shown that
for $T \ll T_K$, the conductance $G(V,B)$ as a function of $V$ and magnetic
field $B$, as well as the magnetization are universal functions of 
$eV/(k_B T_K)$ and $\mu_B B /(k_B T_K)$ for max$(eV,\mu_B B \gg k_B T_K$ \cite{rosch}. In the
opposite limit of small $V$ and $T$, Oguri \cite{ogu1,ogu2} has determined
the scaling of $G(V,T)$ for $B=0$ up to second order in $T$ and $V$ for the
symmetric IAM (SIAM) in which $E_{d}=-U/2$ using a Fermi liquid approach,
extending to finite $V$ the renormalized perturbation theory (RPT) in $U$
developed by Hewson \cite{hew} and using Ward identities. The result is
given as an exact analytical expression in terms of renormalized parameters 
$\widetilde{\Delta} \sim k_B T_K$ and $\widetilde{U}$.

More recent experimental studies for the scaling properties of $G(V,T)$ for small $V$
and $T$ \cite{grobis,scott} stimulated further theoretical work on the
subject \cite{rinc,rati,roura,bal,sela,sca,mu,merker,cb,ng,jpcs,hanl,oh} using
different approximations, like RPT \cite{rinc}, $1/N $ expansion \cite{rati},
non-crossing approximation \cite{roura}, or decoupling of equations of motion \cite{bal}. 
While the results for the Kondo model have been extended to SU(N) symmetry \cite{hanl}, 
to fit experiment, calculations need to include
some degree of valence fluctuations \cite{rinc,roura} suggesting that one has to go
beyond the Kondo model and use the IAM for a quantitative description. The effect of asymmetric
coupling to the left and right leads and asymmetric drop in the bias voltage
has been calculated up to second order in $T$ and $V$ using Fermi liquid
approaches, for the SIAM \cite{rinc,sela,sca}. The more general expression
was given first by Sela and Malecki \cite{sela} and reproduced using RPT \cite{sca}.
These results are exact up to terms of total second order in $V$
and $T$. 

Some of these results of RPT were extended for $E_{d}\neq -U/2$ using two
different approaches. One of them starts from renormalized parameters of the
IAM, $\widetilde{\Delta }$, $\widetilde{U}$ and $\widetilde{E}_{d}$ \cite{sca,ng}. 
The other starts from $\widetilde{\Delta }$ and $\widetilde{U}$ for the symmetric case 
$E_{d}=-U/2$ (for which $\widetilde{E}_{d}=0$) and performs
another perturbation expansion around this point \cite{mu}. The authors call
this approach renormalized superperturbation theory (rSPT) \cite{jpcs}. A controversy
between the authors of both approaches exist for finite voltage $V$ \cite{ng,jpcs,com3},
but this does not affect equilibrium properties
($V=0$) like the one discussed here. In the last years also higher-order Fermi-liquid 
corrections away from half filling were calculated \cite{oh}. 

In Ref. \cite{mu} an analytical expression using rSTP was obtained for the coefficient $c_{T}$ for
the expansion for small $T$ and $V=0$ of the condunctance:
$G(T)=G(0)[1-c_{T}(T/T_{K})^{2}]$. 
This result was compared with a fit of $G(T)$ for small $T$ obtained
using the numerical renormalization group (NRG) \cite{merker}. The comparison
was poor, and rapidly deteriorates with increasing $U$. 
For $U=3\Delta $ the
rSPT expression for $c_{T}$ increases as $E_{d}$ increases from the symmetric
point $E_{d}=-U/2$, while the NRG result decreases.  
Later the authors
included ladder diagrams in their rSPT approach (at the cost of losing an
analytical expression) obtaining a considerable improvement \cite{jpcs}.
However still for $U=3\Delta $ and $E_{d}>-0.6\Delta $ the comparison is
rather poor. Furthermore, the fact that the rSPT results presented are
limited to $U\leq \Delta $ is a shortcoming for two reasons. First, the
Kondo regime $-E_{d},U+E_{d}\gg \Delta $ is not reached. In the Kondo
regime, the spectral density at the dot has in addition to the Kondo peak
near the Fermi energy $\epsilon _{F}$, two charge-transfer peaks at energies
near $E_{d}-\epsilon _{F}$ and $E_{d}+U-\epsilon _{F}$ of total 
width $4\Delta $ \cite{pru,logan,capac,anchos}. 
Then for $U=3\Delta $ even in the symmetric case, the
Kondo peak is merged with one charge-transfer peak and valence fluctuations
are important. Second, for small $U$, simple ordinary (not renormalized)
perturbation theory up to second order $U$ \cite{yos,hz,hor} has been
successful for different problems \cite{levy,kk,nov}. In particular, self-consistent
interpolation schemes \cite{levy,kk,none} permit to extend the validity of the
results for $U$ as large as a few $\Delta$ depending on the problem. 
The interpolative perturbation approach (IPA) 
proposed in Ref. \cite{kk}, and extended to finite magnetic field in Ref. \cite{none} 
has been applied to the coefficient $c_B$ of the expansion of the conductance
with magnetic field $G(B)=G(0)[1-c_{B}(g\mu_B B/k_B T_{K})^{2}]$ \cite{cb,com3}
leading results superior to those of the rSPT including ladder diagrams \cite{com3}.
This IPA requires to satisfy selfconsistently the Friedel sum rule \cite{lan}
to each spin \cite{lady,none}. While this rule cannot be extended to finite temperature $T$,
we have explored a simple extension for $T \rightarrow 0$, as explained in Section 3.

In this work, using RPT we derive an analytical (although lengthy) expression of 
the coefficient $c_T$ for the temperature expansion of the conductance, 
in terms of the renormalized parameters of the model $\widetilde{\Delta }$, $\widetilde{E}_{d}$  
and $\widetilde{U}$. Taking tabulated values of these parameters from the literature
for several values of the original parameters of the model, we obtain the corresponding $c_T$ 
and compare them with published NRG results. The agreement is excellent for most of the calculated points.
The results can be rather easily extended for other sets of parameters in comparison
with NRG for dynamical quantities. We also calculated $c_T$ within the IPA 
and compared with the other approaches.

In Section \ref{forma}, we explain briefly the RPT for the calculation 
of the Green functions and in particular the spectral density of impurity states 
and its expansion for $V=B=0$ and small $\omega$ and $T$.
The expression of $c_T$ is given in Section \ref{cond}. The comparison
with NRG and IPA results is presented in Section \ref{compa}. 
Section \ref{sum}
contains a discussion.

\section{Formalism}

\label{forma}

\subsection{Hamiltonian}

\label{ham}

In the most general case, the model describes a QD interacting with two conducting leads,
one at the left and one at the right, with chemical potentials $\mu _{L}$
and $\mu _{R}$ respectively, with $\mu _{L}$-$\mu _{R}=eV$. 
The system is at temperature $T$ in presence of a magnetic field $B$.
For the sake of completeness we begin discussing the general case,
and later we take $V=B=0$.
The dot level
has an on-site energy $E_{d}$ controlled by a gate voltage and an on-site
repulsion $U.$ 
The Hamiltonian is that of the IAM 

\begin{eqnarray}
H &=&\sum_{k\nu \sigma }\varepsilon _{k\nu }c_{k\nu \sigma }^{\dagger
}c_{k\nu \sigma }+\sum_{\sigma }E_{d}^{\sigma }n_{d\sigma }
 \nonumber
\\
&&+\sum_{k\nu \sigma }(V_{k\nu }c_{k\nu \sigma }^{\dagger }d_{\sigma }+{\rm H.c.}) 
+Un_{d\uparrow }n_{d\downarrow }.  \label{h}
\end{eqnarray}
Here $\nu =L,R$ refers to the left and right leads and the operator $c_{k\nu
\sigma }^{\dagger }$ creates an electron in the state with wave vector $k$
and spin $\sigma $ at the lead $\nu $, Similarly $d_{\sigma }^{\dagger }$
creates an electron with spin $\sigma $ at the QD.  The number operator $%
n_{d\sigma }=d_{\sigma }^{\dagger }d_{\sigma }$ and $E_{d}^{\sigma
}=E_{d}-\sigma \mu _{B}B$. We assume coupling to the leads $\Delta _{\nu
}=\pi \sum_{k}|V_{k\nu }|^{2}\delta (\omega -\varepsilon _{k\nu })=\beta
_{\nu }\Delta $ independent of energy, and define the total resonant level
width $\Delta =\Delta _{L}+\Delta _{R}$.

\subsection{Green function within renormalized perturbation theory}

\label{green}

For a symmetric flat band of conduction states and constant $\Delta $ as we
have assumed, the retarded Green function of the QD level for spin $\sigma $
can in general be written as

\begin{equation}
G_{d\sigma }(\omega ) = \frac{1}{\omega -E_{d}^{\sigma }+i\Delta -\Sigma
_{\sigma }(\omega )},  \label{gr}
\end{equation}%
where $\Sigma _{\sigma }(\omega )$ is the (unknown) retarded self energy.

The basic idea of RPT is to reorganize the perturbation expansion in terms
of fully dressed quasiparticles in a Fermi liquid picture \cite{hew}. The
parameters of the original model are renormalized and the renormalized 
values $\widetilde{\Delta }$, $\widetilde{U}$ and $\widetilde{E}_{d}^{\sigma
}$ can be calculated exactly from Bethe ansatz results \cite{tsve,tsve2,bethe}, 
or accurately using
NRG. One of the main advantages is that the renormalized expansion parameter 
$u=\widetilde{U}/(\pi \widetilde{\Delta })$ is small. 
In general $u\lesssim 1$. In the following we set the origin of
one-particle energies at the Fermi level ($\epsilon _{F}=0$). Within RPT,
the low-energy part of $G_{d\sigma }(\omega )$ is approximated expanding the
denominator around $\omega =\epsilon _{F}=0$, for $V=T=0.$\cite{hew,ng}

\begin{equation}
G_{d\sigma }(\omega )\simeq \frac{z}{\omega -\widetilde{E}_{d}^{\sigma }+i%
\widetilde{\Delta }-\widetilde{\Sigma }_{\sigma }(\omega )},  \label{gra}
\end{equation}%
where $z=[1-\partial \Sigma _{\sigma }(\omega )/\partial \omega |_{\omega
=0}]^{-1}$ is the quasiparticle weight, $\widetilde{\Delta }=z\Delta $ is
the renormalized resonant level width,  $\widetilde{E}_{d}=z[E_{d}+\Sigma
_{\sigma }(0)]$ is the renormalized level energy and

\begin{equation}
\widetilde{\Sigma }_{\sigma }(\omega )=z[\Sigma _{\sigma }(\omega )-\Sigma
_{\sigma }(0)-\omega \partial \Sigma _{\sigma }(\omega )/\partial \omega
|_{\omega =0}].  \label{sigren}
\end{equation}%
We emphasize that $\Sigma _{\sigma }(0)$ and $\partial \Sigma _{\sigma
}(\omega )/\partial \omega |_{\omega =0}$ are calculated at $V=T=\omega =0$.%
\cite{com3}

The renormalized Coulomb repulsion $\widetilde{U}$ is given by a vertex
function \cite{ogu1,ogu2,hew}.

The spectral density of $d$ electrons is 

\begin{equation}
\rho _{\sigma }(\omega )=-{\rm Im}G_{d\sigma }(\omega )/\pi .  \label{rho}
\end{equation}

The free quasiparticle spectral density of $d$ electrons is given by 
\begin{equation}
\widetilde{\rho }_{0}^{\sigma }(\omega )=\frac{\widetilde{\Delta }/\pi }{%
(\omega -\widetilde{E}_{d}^{\sigma })^{2}+\widetilde{\Delta }^{2}}.
\label{rhor}
\end{equation}%
Using Friedel sum rule \cite{lan,lady} one has

\begin{equation}
\pi \Delta \rho _{\sigma }(0)=\pi \widetilde{\Delta }
\widetilde{\rho }_{0}^{\sigma }(0)=\sin ^{2}(\pi \langle n_{d\sigma }\rangle ).  \label{rho0}
\end{equation}%
Thus, knowing the occupancies $\langle n_{d\sigma }\rangle $ experimentally
or by a Bette ansatz calculation for example, one can determine the ratios $%
\widetilde{E}_{d}^{\sigma }/\widetilde{\Delta }=\cot (\pi \langle n_{d\sigma
}\rangle )$. The ratio $\widetilde{U}/\widetilde{\Delta }$ can be obtained
from the expression of the impurity contribution to magnetic susceptibility
at zero temperature \cite{hew}

\begin{equation}
\chi =(g\mu _{B})^{2}\widetilde{\rho }_{0}(0)(1+\widetilde{U}\widetilde{\rho 
}_{0}(0))/2,  \label{xi}
\end{equation}
where $\widetilde{\rho }_{0}(\omega)=\widetilde{\rho }_{0}^{\uparrow }(\omega)
=\widetilde{\rho }_{0}^{\downarrow }(\omega)$ for $B=0$
and $\widetilde{\Delta }$ can be obtained either from the linear term $\gamma _{C}$ in the
impurity contribution to the specific heat \cite{hew}

\begin{equation}
\widetilde{\Delta }=\frac{2\pi k_{B}^{2}}{3\gamma _{C}}\sum_{\sigma }\sin
^{2}(\pi \langle n_{d\sigma }\rangle ),  \label{delt}
\end{equation}%
or approximately in RPT from the half-width at half maximum of the Kondo
peak in $\rho _{\sigma }(\omega )$ \cite{cb}.

To obtain the spectral density $\rho _{\sigma }(\omega )$ out of the point $%
\omega =T=V=0$, we need an approximation for $\widetilde{\Sigma }_{\sigma
}(\omega )$. As in previous works \cite{rinc,cb,ng} we use

\begin{equation}
\widetilde{\Sigma }_{\sigma }(\omega )=\widetilde{\Sigma }_{\sigma
}^{2}(\omega )-\widetilde{\Sigma }_{\sigma }^{2}(0)-\omega \partial 
\widetilde{\Sigma }_{\sigma }^{2}/\partial \omega |_{\omega =0},
\label{remre}
\end{equation}%
where $\widetilde{\Sigma }_{\sigma }^{2}(\omega )$ is obtained using
perturbation theory up to second order in $\widetilde{U}$, using the free
quasiparticle spectral density $\widetilde{\rho }_{0}^{\sigma }(\omega )$
[or the corresponding Green function $1/(\omega -\widetilde{E}_{d}^{\sigma
}+i\widetilde{\Delta })$]. Since the constant first-order term vanishes in
Eq. (\ref{remre}), a possible expression for $\widetilde{\Sigma }_{\sigma
}^{2}(\omega )$ is \cite{none}

\begin{eqnarray}
\Sigma _{\uparrow }^{2}(\omega ) &=&\widetilde{U}^{2}\int d\epsilon
_{1}d\epsilon _{2}d\epsilon _{3}\frac{\widetilde{\rho }_{0}^{\uparrow
}(\epsilon _{1})\widetilde{\rho }_{0}^{\downarrow }(\epsilon _{2})\widetilde{%
\rho }_{0}^{\downarrow }(\epsilon _{3})}{\omega +\epsilon _{3}-\epsilon
_{1}-\epsilon _{2}+i\eta }  \nonumber \\
&&\times \lbrack (1-\tilde{f}(\epsilon _{1}))(1-\tilde{f}(\epsilon _{2}))%
\tilde{f}(\epsilon _{3})  \nonumber \\
&&+\tilde{f}(\epsilon _{1})\tilde{f}(\epsilon _{2})(1-\tilde{f}(\epsilon
_{3}))],  \label{srro}
\end{eqnarray}%
where $\tilde{f}(\omega )=\sum_{\nu }\beta _{\nu }f(\omega -\mu _{\nu })$,
with $f(\omega )$ the Fermi function,  and the same interchanging spin up
and down.

The lesser and greater Green functions are defined similarly \cite{ng}.

\subsection{Expansion of the renormalized retarded self energy}

\label{exp}

In the following, we take $V=B=0.$ Then $\widetilde{E}_{d}^{\sigma }=\widetilde{E}_{d}$ 
independent of $\sigma$.
Borrowing results from Ref. \cite{hz} for
the expansion of $\widetilde{\Sigma }_{\sigma }^{2}(\omega )$ up to total
second order in $\omega $ and $T$,  and inserting them in Eq. (\ref{remre})
we obtain 

\begin{equation}
\widetilde{\Sigma }_{\sigma }(\omega )=-u^{2}\frac{\alpha \omega ^{2}+\beta (\pi
k_{B}T)^{2}+i\gamma \lbrack \omega ^{2}+(\pi k_{B}T)^{2}]}{\widetilde{\Delta 
}},  \label{sigexp}
\end{equation}%
where we define

\begin{eqnarray}
u&=&\frac{\widetilde{U}}{\pi 
\widetilde{\Delta }} {\rm , }  \epsilon  =\frac{\widetilde{E}_{d}}{\widetilde{\Delta }},  \nonumber \\
s &=&\sin (\pi \langle n_{d\sigma }\rangle )=\frac{\widetilde{\Delta }}{%
\sqrt{\widetilde{E}_{d}^{2}+\widetilde{\Delta }^{2}}},  \nonumber \\
c &=&\cos (\pi \langle n_{d\sigma }\rangle ),  \label{auxi}
\end{eqnarray}%
and the coefficients $\alpha $, $\beta $, and $\gamma $ are given by

\begin{eqnarray}
\alpha  &=&s^{4}(t_{1}+t_{2}),  \nonumber \\
t_{1} &=&\frac{\arctan (\epsilon )/\epsilon -s^{2}}{4\epsilon },  \nonumber \\
t_{2} &=&\arctan (\epsilon )[9/4+2\epsilon \arctan (\epsilon )]  \nonumber \\
&&+ \epsilon s^{2}[\frac{13-3\pi ^{2}+(\pi \epsilon )^{2}}{4}  \nonumber \\
&&+(1-3\epsilon ^{2})g(\epsilon )],  \nonumber \\
g(\epsilon ) &=&\frac{1}{\epsilon }\int\limits_{0}^{\epsilon }dt[\arctan^{2}(t)  \nonumber \\
&&+\frac{2}{t}\arctan ^{2}(t)],  \label{alpha}
\end{eqnarray}

\begin{equation}
\beta =\frac{s^{2}}{3}[t_{1}(1+5\epsilon ^{2})-\frac{\epsilon s^{2}}{2}],
\label{beta}
\end{equation}

\begin{equation}
\gamma =\frac{s^{4}}{2}.  \label{gamma}
\end{equation}

\subsection{Conductance as a function of temperature}

\label{cond}

In linear response ($V\rightarrow 0$) and for $B=0$, the conductance is
given by \cite{meir}

\begin{equation}
G(T)=C\int d\omega \rho _{\sigma }(\omega ,T)\left( -\frac{\partial
f(\omega )}{\partial \omega }\right) ,  \label{gt}
\end{equation}%
where $C$ is a constant that depends on the couplings $\Delta _{L}$ and $%
\Delta _{R}$.  

Up to second order in the  temperature $T$, using the Sommerfeld expansion
one has

\begin{equation}
G(T)\simeq C[\rho _{\sigma }(0,T)+\frac{(\pi k_{B}T)^{2}}{6}\frac{\partial
^{2}\rho _{\sigma }(\omega ,0)}{\partial \omega ^{2}}|_{\omega =0}.
\label{gtexp}
\end{equation}%
Using Eqs. (\ref{gra}), (\ref{rho}), (\ref{sigexp}), (\ref{auxi}), (\ref{alpha}), 
(\ref{beta}), (\ref{gamma}), and (\ref{gtexp}) we obtain after
some algebra the desired expression for the leading temperature dependence
of the equilibrium conductance

\begin{eqnarray}
\frac{G(T)}{G(0)} &=&1+\left( \frac{\pi k_{B}T}{\widetilde{\Delta }}\right)^2 [%
\frac{s^{2}}{3}(4c^{2}-1)  \nonumber \\
&&+u^{2}\{2\left( \frac{\alpha }{3}+\beta \right) sc+\frac{4\gamma }{3}(1-2s^{2})\}].
\label{gtf}
\end{eqnarray}

\section{Comparison with NRG for dynamical quantities and IPA}

\label{compa}

\begin{figure}[tbp]
\includegraphics[width=\columnwidth]{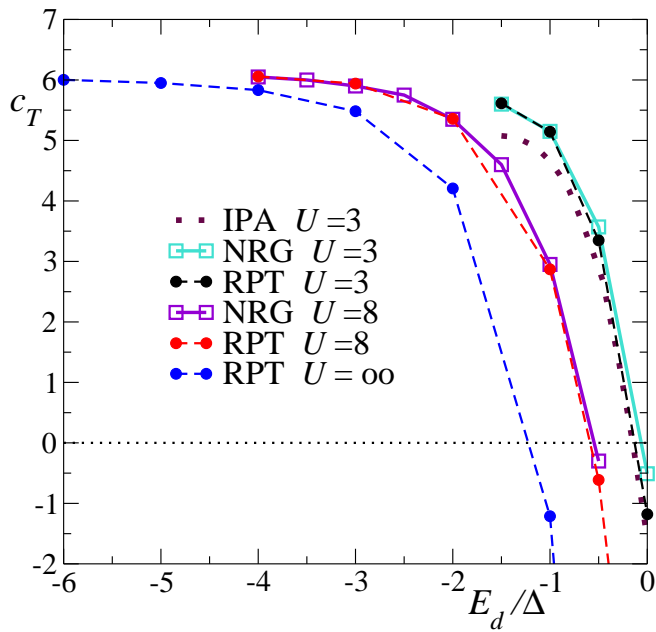}
\caption{Coefficient $c_T$ of Eq. (\ref{ct})  vs 
$E_{d}$ for several values of $U$.}
\label{ctf}
\end{figure}

In Ref. \cite{merker}, the coefficient $c_{T}$ was defined as

\begin{equation}
\frac{G(T)}{G(0)}=1-c_{T}\left( \frac{T}{T_{0}}\right) ^{2},  \label{ct}
\end{equation}%
where  $T_{0}$ is of the order of the Kondo temperature and defined in terms
of the magnetic susceptibility $\chi $ by

\begin{equation}
T_{0}=\frac{\left( g\mu _{B}\right) ^{2}}{4k_{B}\chi }.  \label{t0}
\end{equation}%
Eqs. (\ref{rhor}) and (\ref{xi}) permit to express $T_{0}$ in terms of the
renormalized parameters. 
For $-E_d = U/2 \rightarrow \infty$, $T_0 = \pi \widetilde{\Delta } /(4 k_B)$ \cite{hew}
and $c_T= \pi^4 /16 \approx 6.09$ \cite{ogu1,ogu2}.
The values of the renormalized parameters parameters were calculated in
Ref. \cite{cb} following the procedure explained by Hewson {\it et al.} \cite{hom}. 
They are reproduced in Table \ref{tabu} for the ease of the reader.
The original parameters include $U= 8 \Delta$ (for which the system is in the Kondo regime 
near the symmetric point $E_d=-U/4$), 
and $U \rightarrow + \infty$ which is more realistic for several molecular QDs \cite{serge}.
Using these renormalized parameters, we have
calculated $c_{T}$ using the expression of the previous section. 
The results are shown in Fig. \ref{ctf}.

\begin{table}[h]
\caption{\label{tabu} Renormalized parameters $\widetilde{\Delta}/\Delta$, 
$\epsilon=\widetilde{E}_d/\widetilde{\Delta}$ 
and $u=\widetilde{U}/(\pi \widetilde{\Delta})$ obtained from NRG+RPT 
for several values of $U / \Delta$ and
$E_d / \Delta$ \cite{cb}. }
\begin{tabular}{lllll}
$U / \Delta$&$E_d/\Delta$ & $\widetilde{\Delta}/\Delta$ & $\epsilon$ & 
$u$  \\ \hline
3 &-1.5  & 0.639 & 0     & 0.738  \\
3 &-1    & 0.671 & 0.196 & 0.732  \\
3 &-0.5  & 0.754 & 0.421 & 0.716  \\
3 &0     & 0.845 & 0.700 & 0.698  \\ \hline
8 &-4 & 0.120 & 0     & 0.985  \\
8 &-3 & 0.143 & 0.101 & 0.987  \\
8 &-2 & 0.235 & 0.247 & 1.004  \\
8 &-1 & 0.457 & 0.510 & 1.040  \\
8 &-0.5 & 0.609 & 0.715 & 1.060  \\
8 &0 & 0.746 & 0.977 & 1.081  \\ \hline
$+ \infty$ &-6 & $2.51 \times 10^{-4}$ & 0.0937 & 1.009  \\
$+ \infty$ &-5 & $1.21 \times 10^{-3}$ & 0.118 & 1.014  \\
$+ \infty$ &-4 & $5.79 \times 10^{-3}$ & 0.160 & 1.025  \\
$+ \infty$ &-3 & 0.0270 & 0.243 & 1.054  \\
$+ \infty$ &-2 & 0.115 & 0.416 & 1.136  \\
$+ \infty$ &-1 & 0.356 & 0.766 & 1.317  \\
$+ \infty$ &0 & 0.640 & 1.338 & 1.594  \\
\end{tabular}
\end{table}

We also show $c_T$ for the same values of $U$ as those in Table \ref{tabu} 
reported in Fig. 5 of Ref. \cite{merker}. 
In that work, $c_T$ has been extracted from a fit to Eq. (\ref{ct}) of several 
low-temperature values (in the range $10^{-5}T_0 \le T \le 0.02 T_0$) of the conductance $G(T)$
obtained using an NRG for dynamical quantities developed in Ref. \cite{yso}. 
In addition two values for $z$-averaging were used \cite{merker}.

For a moderate value of $U= 3 \Delta$, we also show the results of the IPA. 
These were obtained with the following procedure. First we used for the self-energy 
the result based on second-order perturbation theory at $T=V=0$ and finite magnetic field $B$,
as in Refs. \cite{none}. The unperturbed Green functions  

\begin{equation}
G^0_{d\sigma }(\omega ) = \frac{1}{\omega - \varepsilon_{d}^{\sigma }+i\Delta},  \label{gun}
\end{equation}
corresponding to the unperturbed Hamiltonian 
\begin{equation}
H_0=H-\sum_{\sigma }(E_{d}^{\sigma }-\varepsilon_{d}^{\sigma })n_{d\sigma } -Un_{d\uparrow }n_{d\downarrow },
\end{equation}
are calculated with effective on-site energies 
$\varepsilon_{d}^{\sigma }$ determined self consistently to satisfy the Friedel sum rule for both spins \cite{lady}.
From this calculations we extract $\varepsilon_{d}=\varepsilon_{d}^{\uparrow }=\varepsilon_{d}^{\downarrow }$ for $B=0$ and 
the magnetic susceptibility from numerical differentiation of the magnetization. Using Eq. (\ref{t0}) $T_0$
is obtained. Then, we calculate the IPA self-energy at finite temperature keeping $\varepsilon_{d}$ fixed, 
and fit the low-$T$ results to a quadratic dependence.

Since $c_T$ has the same value replacing $E_d$ by $U-E_d$ we represent in Fig. \ref{ctf} only $E_d \ge -U/2$. 
It is apparent that $c_T$ decreases monotonically showing a downward curvature
with increasing (or decreasing) $E_d$  starting from the symmetric point $E_d=-U/2$,
becoming negative for $E_d \sim 0$.

It is clear that the comparison between RPT and NRG results are very good. 
For positive $c_T$ the difference is of the order of the symbol size
and increases as the on-site energy $E_d$ is moved away from the symmetric point.
For $U= 3 \Delta$ the maximum difference between the values included in the figure is 0.67 for $E_d=0$
(12 \% of the maximum value $c_T=5.61$ for $E_d=-U/2$).

In Ref. \cite{merker} also the coefficient $c_T^\prime$ was introduced which differs from $c_T$ in
the fact that the characteristic temperature $T_0$ was taken always as that of the symmetric point
$T_0^{\rm  sym} \le T_0$. The relation between both coefficients is

\begin{equation}
\frac{c_T^\prime}{c_T}=\left(\frac{T_0^{\rm  sym}}{T_0}\right)^2.  
\label{ctp}
\end{equation}

The above mentioned difference in $c_T$ is reduced by a factor
0.18 ($k_B T_0=1.3468 \Delta$, $k_B T_0^{\rm  sym}=0.5775 \Delta$) in $c^\prime_T$. Then, the maximum deviation 
in $c^\prime_T$ for $U= 3 \Delta$ is below 0.1. 
In Ref. \cite{jpcs}, a comparison between result of $c^\prime_T$ calculated with NRG and
rSPT including ladder diagrams was presented for $U \leq 3 \Delta$. From Fig. 1 of Ref. \cite{jpcs}, 
it is clear that the deviation of 
both results is already larger than 0.8 for $U= 3 \Delta$ and 
$E_d = -0.3 \Delta$. This indicates that our RPT results for $U = 3 \Delta$ are nearly an order of magnitude 
more precise near $E_d=0$. Note that the rSPT results depend on two parameters, $\widetilde{\Delta }$, $\widetilde{U}$
for $E_d=-U/2$, while in our RPT approach one has in addition $\widetilde{E}_{d}$ and all parameters depend on $E_d$.

For $U= 3 \Delta$ we also show the results obtained using the IPA. In contrast to RPT and rSPT, the results
do not depend on renormalized parameters. As a consequence, while RPT and rSPT give by construction the exact
result at the symmetric point $E_d=-U/2$ taking known values of $\widetilde{\Delta }$ and $\widetilde{U}$ with
$\widetilde{E}_{d}=0$, IPA deviates from the correct result. This is due to its inaccuracy in the calculation of 
the magnetic susceptibility (which determines the energy scale $T_0$), underestimated by 8 \% and also an
underestimation of the curvature of $G(T)$. The accuracy of the IPA increases away from the symmetric point, 
and taking into accounts its simplicity, the IPA provides a rather good semiquantitative description for 
$ U = 3 \Delta$ (or lower), 
although $c_T$ continues underestimated in the whole range of $E_d$. 
The IPA seems to be better than the rSPT near the intermediate valence region.

The comparison between RPT and NRG for $U=8 \Delta$ shows that the agreement does not deteriorate with increasing 
$U$ in contrast to the case of IPA \cite{cb} or rSPT \cite{merker,jpcs,com3}. For example, the underestimation 
of the magnetic susceptibility at the symmetric point by the IPA increases to 15 \% for $U= 4 \Delta$, while 
it is only 1.4 \% for $U= 2 \Delta$.

\section{Summary and discussion}

\label{sum}

Using renormalized perturbation theory (RPT), we have provided an analytical expression for the coefficient of the 
leading temperature dependence of the conductance through a quantum dot, in terms of the renormalized parameters 
of the impurity Anderson model $\widetilde{\Delta }$, $\widetilde{E}_{d}$  
and $\widetilde{U}$. The expression is given by Eq. (\ref{gtf}) where the different coefficients
are defined by Eqs. (\ref{auxi}), 
(\ref{alpha}), (\ref{beta}) and (\ref{gamma}). 
Using Eqs. (\ref{ct}) and (\ref{t0}) the coefficient $c_T$ defined previously \cite{merker} is immediately obtained, 
and also $c^\prime_T$ [see Eq. (\ref{ctp})] which uses a fixed Kondo scale evaluated at $E_d=-U/2$.
Although the expression is lengthy it can be easily evaluated. The most difficult task is a
one-dimensional integration [last Eq. (\ref{alpha})]. 
The renormalized parameters can be easily obtained from the
spectrum of an NRG calculation \cite{hom} or from the calculation of static quantities 
with Bethe ansatz \cite{tsve,tsve2,bethe}, or from experiment.
Refs. \cite{tsve,tsve2} provide analytical expressions for the occupancy, magnetic susceptibility 
and specific heat, from which the renormalized parameters can be calculated using Eqs. (\ref{rho}) to (\ref{delt}).
Some tricks to evaluate integrals that enter these expressions are given in the appendix of Ref. \cite{anda}.

The calculation of dynamical quantities like the conductance
is not possible with Bethe ansatz, and much more complicated within NRG \cite{hof,bulla}.
To calculate $c_T$ directly within NRG for dynamical properties required several calculations
at different temperatures within an optimized range of temperatures and for
two different logarithmic discretizations ($z$-averaging)  \cite{merker}.
We would like to notice that even the calculation of the static magnetic susceptibility $\chi$ 
[which determines $T_0$, see Eq. (\ref{t0})]  within NRG is 
much easier determining first the renormalized parameters and then using Eq. (\ref{xi}), as done in Ref. \cite{cb}. 
A direct calculation of $\chi$ using standard NRG well inside the Kondo regime displays oscillations with 
temperature and even negative values \cite{fang,wong}. A full density-matrix NRG was required to solve this 
problem \cite{fang}, but this is not necessary to calculate the renormalized parameters \cite{hom}.

Our calculation with RPT is therefore much easier than direct evaluation of the conductance using NRG. 
It also has the advantage over ordinary (not renormalized) perturbation approaches \cite{cb,com3},
or the so called  renormalized superperturbation theory (rSPT) including ladder diagrams \cite{jpcs}
that the results do not deteriorate rapidly with increasing $U$ allowing us to reach the Kondo regime 
$-E_d, U+E_d \gg \Delta$.

Different approaches discussed here (RPT, rSPT, NRG, but not IPA) give the same correct value of $c_T$ at the symmetric point $E_d=-U/2$.
Also by definition, $c_T$ and $c^\prime_T$ coincide at this point [see Eq. (\ref{ctp})] Out of this point, since 
$T_0$ can be considerably larger than $T_0^{\rm  sym}$, the magnitude of $c^\prime_T$ is smaller or much smaller than
$c_T$. 
From this analysis, it is clear that plotting $c_T$ instead of $c^\prime_T$ is more appropriate to see differences 
between different approaches. Moreover $T_0$ is directly related with the width of the Kondo peak in the spectral 
density of states $2\Delta_\rho$ (the ratio $\Delta_\rho/T_0$ has been calculated within RPT in Ref. \cite{cb}),
which in turn is of the order of the width $2\Delta_G$ of the zero-bias anomaly in the conductance 
$G(V)$ \cite{capac},
which is experimentally accessible. The ratio $\Delta_G/\Delta_\rho$ depends on the ratio of the couplings between 
left and right leads $\Delta_L/\Delta_R$ and has been calculated \cite{capac}.

In the Kondo regime, an empirical formula that fits very well the NRG results for the temperature 
dependence of the conductance has been proposed. \cite{gg}. It can be written in the form

\begin{equation}
G(T)=\frac{G(0)}{\left[ 1+(2^{1/s}-1)(T/T^G_{K})^{2}\right] ^{s}},
\label{ge}
\end{equation}
where $s=0.22$ and $T^G_{K}$ (of the order of $T_0$) is the temperature at which the conductance falls
to half of the zero temperature value: $G(T^G_{K})=G(0)/2$. One may wonder to which extent the expansion of 
this expression for $T \rightarrow 0$

\begin{eqnarray}
\frac{G(T)}{G(0)} & \approx & 1-c_E \left( \frac{T}{T^G_{K}}\right) ^{2} \nonumber \\
c_E &=& s(2^{1/s}-1) \approx 4.92, 
\label{ce}
\end{eqnarray}
gives the correct $c_T= \pi^4 /16 \approx 6.09$ in the Kondo limit. Comparing Eqs. (\ref{ct}) and (\ref{ce}),
one realizes that to answer this question one needs to know the ratio $T^G_{K}/T_0$. 
We have calculated this ratio for two cases presented above: $U \rightarrow \infty$, $E_d/\Delta= -6$,
and $U/\Delta=8 $, $E_d/\Delta= -4$. For the first case, Eqs. (\ref{rhor}), (\ref{xi}) 
and the data of Table \ref{tabu} give  $\chi=0.99/(\pi \widetilde{\Delta})$ 
and then from Eq. (\ref{t0}) 
$k_B T_0  \approx 0.79 \widetilde{\Delta }$. 
Taking $\widetilde{\Delta}=2.51 \times 10^{-4} \Delta$ from
Table \ref{tabu}, one obtains $k_B T_0 = 1.99 \times 10^{-4} \Delta$, which almost coincides with the value 
$k_B T^G_{K} = 1.98 \times 10^{-4} \Delta$ obtained using NRG for dynamical quantities \cite{alejo}.
Since the value obtained by RPT for these parameters is $c_T=6.00$ (see Fig. \ref{ctf}),
$c_E$ is an underestimation by 17 \%. It is interesting to note that the RPT calculation of
the half width at half maximum of the spectral density for these parameters is \cite{cb}
$\Delta_{\rho}=0.706 \widetilde{\Delta}=0.89 k_B T_0$.  
Similarly, for $U/\Delta=8 $, $E_d/\Delta= -4$, we obtain $c_T=6.05$, $\Delta_{\rho} = 0.90 k_B T_0$ 
$k_B T_0 = 0.791 \widetilde{\Delta}=0.095 \Delta$, while $k_B T^G_{K} = 0.101 \Delta$ \cite{alejo}.     
In this case, $c_E/c_T=0.72$.
The failure of Eq. (\ref{ge}) to accurately reproduce the low-$T$ behavior in the Kondo regime 
is due to the fact that it was 
devised to fit the conductance in a wide temperature range and not just for small $T$.

\section*{Acknowledgments}

We are indebted to J. A. Andrade for his  NRG calculations of $T^G_{K}$.
This work was sponsored by PIP 112-201101-00832 of CONICET and PICT
2013-1045 of the ANPCyT.

\end{document}